\documentclass[11pt]{cernrep}
\usepackage{graphicx}
\usepackage{cite,./mcite}
\usepackage{xspace}
\usepackage[latin1]{inputenc}

\newcommand{\pythia}{P\scalebox{0.8}{YTHIA}\xspace}

\newcommand{\amegic}{\scalebox{0.8}{AMEGIC++}\xspace}
\newcommand{\apacic}{\scalebox{0.8}{APACIC++}\xspace}
\newcommand{\herwig}{\scalebox{0.8}{HERWIG}\xspace}
\newcommand{\ariadne}{A\scalebox{0.8}{RIADNE}\xspace}
\newcommand{\sherpa}{S\scalebox{0.8}{HERPA}\xspace}
\newcommand{\alpgen}{A\scalebox{0.8}{LPGEN}\xspace}

\begin{document}

\title{Matching Parton Showers and Matrix Elements}

\author{Stefan H\"oche$^1$, Frank~Krauss$^1$, Nils~Lavesson$^2$,
   Leif~L\"onnblad$^2$, Michelangelo~Mangano$^3$,
   Andreas~Sch\"alicke$^1$, Steffen~Schumann$^1$}

\institute{$^1$Institut f\"ur Theoretische Physik, TU Dresden,
   Germany; $^2$Department of Theoretical Physics, Lund University,
   Sweden; $^3$CERN, Geneva, Switzerland.}

\maketitle
\begin{abstract}
   We compare different procedures for combining fixed-order tree-level
   matrix element generators with parton showers. We use the case of
   W-production at the Tevatron and the LHC to compare different
   implementations of the so-called CKKW scheme and one based on the
   so-called MLM scheme using different matrix element generators and
   different parton cascades. We find that although similar results are
   obtained in all cases, there are important differences.
\end{abstract}

\section{Introduction}
One of the most striking features of LHC final states will be the
large number of events with several hard jets. Final states with 6
jets from $t\bar{t}$ decays will have a rate of almost 1Hz, with
10-100 times more coming from prompt QCD processes. The immense amount
of available phase-space, and the large acceptance of the detectors,
with calorimeters covering a region of almost 10 units of
pseudorapidity ($\eta$), will lead to production and identification of
final states with 10 or more jets. These events will hide or strongly
modify all possible signals of new physics which involve the chain
decay of heavy coloured particles, such as squarks, gluinos or the
heavier partners of the top which appear in little-Higgs models. Being
able to predict their features is therefore essential.

To achieve this, our calculations need to describe as accurately as
possible both the full matrix elements for the underlying hard
processes, as well as the subsequent development of the hard partons
into jets of hadrons. For the complex final-state topologies we are
interested in, no factorization theorem exists however to rigorously
separate these two components, providing a constructive algorithm for
the implementation of such separation. The main obstacle is the
existence of several hard scales, like the jet transverse energies and
dijet invariant masses, which for a generic multijet event will span a
wide range. This makes it difficult to unambiguously separate the
components of the event which belong to the ``hard process'' (to be
calculated using a multiparton amplitude) from those developing during
its evolution (described by the parton shower). A given $(N+1)$-jet
event can be obtained in two ways: from the collinear/soft-radiation
evolution of an appropriate $(N+1)$-parton final state, or from an
$N$-parton configuration where hard, large-angle emission during its
evolution leads to the extra jet.  A factorization prescription (in
this context this is often called a ``matching scheme'') defines, on
an event-by-event basis, which of the two paths should be followed.
The primary goal of a matching scheme is therefore to avoid double
counting (by preventing some events to appear twice, once for each
path), as well as dead regions (by ensuring that each configuration is
generated by at least one of the allowed paths). Furthermore, a good
matching scheme will optimize the choice of the path, using the one
which guarantees the best possible approximation to a given
kinematics. It is possible to consider therefore different matching
schemes, all avoiding the double counting and dead regions, but
leading to different results in view of the different ways the
calculation is distributed between the matrix element and the shower
evolution. As in any factorization scheme, the physics is independent
of the separation between phases only if we have complete control over
the perturbative expansion. Otherwise a residual scheme-dependence is
left. Exploring different matching schemes is therefore crucial to
assess the systematic uncertainties of multijet calculations.

In this work we present a first comparison of the three approaches
which have been proposed so far, the so-called CKKW scheme, the
L\"onnblad scheme, and the MLM scheme. After shortly reviewing them,
we present predictions for a set of $W$+multijet distributions at the
Tevatron collider and at the LHC.

\section{Matching procedures}
\label{sec:matching-procedures}

\noindent
In general, the different merging procedures all follow a similar  
strategy:
\begin{enumerate}
\item A jet measure is defined and all relevant cross sections
  including jets are calculated for the process under consideration.
  I.e.\ for the production of a final state $X$ in $pp$-collisions,
  the cross sections for the processes $pp\to X+n\mbox{\rm jets}$ with
  $n=0,\,1,\,\dots n_{\rm max}$ are evaluated.
\item Hard parton samples are produced with a probability proportional
  to the respective total cross section, in a corresponding kinematic
  configuration following the matrix element.
\item The individual configurations are accepted or rejected with a
  dynamical, kinematics-dependent probability that includes both
  effects of running coupling constants and of Sudakov effects. In
  case the event is rejected, step 2 is repeated, i.e.\ a new parton
  sample is selected, possibly with a new number of jets.
\item The parton shower is invoked with suitable initial conditions
  for each of the legs. In some cases, like, e.g.\ in the MLM
  procedure described below, this step is performed together with the
  step before, i.e.\ the acceptance/rejection of the jet
  configuration. In all cases the parton shower is constrained not to
  produce any extra jet; stated in other words: Configurations that
  would fall into the realm of matrix elements with a higher jet
  multiplicity are vetoed in the parton shower step.
\end{enumerate}
From the description above it is clear that the merging procedures
discussed in this contribution differ mainly
\begin{itemize}
\item in the jet definition used in the matrix elements;
\item in the way the acceptance/rejection of jet configurations
  stemming from the matrix element is performed;
\item and in details concerning the starting conditions of and the jet
  vetoing inside the parton showering.
\end{itemize}

\subsection{CKKW}
\label{sec:ckkw}
\noindent
In the original merging description according to
\cite{Catani:2001cc,Krauss:2002up}, which has been implemented
\cite{Schalicke:2005nv} in \sherpa \cite{Gleisberg:2003xi} in full
generality, the acceptance/rejection of jet configurations from the
matrix elements and the parton showering step are well-separated.

\noindent
In this realisation of what is known as the CKKW-prescription the
phase space separation for the different multijet processes is
achieved through a $k_\perp$-measure
\cite{Catani:1991hj,Catani:1992zp,Catani:1993hr}. For the case of
hadron--hadron collisions, two final-state particles belong to two
different jets, if their relative transverse momentum
\begin{eqnarray}
   k_\perp^{(ij)2} = 2\,\mbox{\rm min}\left\{p_\perp^{(i)},
     \,p_\perp^{(j)}\right\}^2
   \left[\cosh(\eta^{(i)}-\eta^{(j)})-\cos(\phi^{(i)}-\phi^{(j)})\right]
\end{eqnarray}
is larger than a critical value, $k^2_{\perp,0}$. In addition, the
transverse momentum of each jet has to be larger than $k_{\perp,0}$.
The matrix elements are then reweighted by appropriate Sudakov and
coupling weights. The task of the weight attached to a matrix element
is to take into account terms that would appear in a corresponding
parton shower evolution. Therefore, a ``shower history'' is
reconstructed by clustering the initial and final state partons
according to the $k_\perp$-algorithm. The resulting chain of nodal
$k_\perp$-measures is interpreted as the sequence of relative
transverse momenta of multiple jet production. The first ingredient of
the weight are the strong coupling constants taken at the respective
nodal values, divided by the value of $\alpha_S$ used during the
matrix element evaluation. The other part of the correction weight is
provided by NLL-Sudakov form factors defined by

\begin{equation}
   \Delta_{q,g}(Q,Q_0) :=
   \exp\left[-\int\limits_{Q_0}^Q\mbox{\rm d}q\Gamma_{q,g}(Q,q)\right] 
\,,
   \label{eq:sud}
\end{equation}
where the integrated splitting functions $\Gamma_{q,g}$ are given by
\begin{equation}
   \Gamma_{q,g}(Q,q) := \left\{\begin{array}{l}
       \frac{2C_F\alpha_s(q)}{\pi q}\left[\log\frac{Q}{q}-\frac{3}{4} 
\right]\\
       \frac{2C_A\alpha_s(q)}{\pi q}\left[\log\frac{Q}{q}-\frac{11} 
{12}\right]
     \end{array}\right.
   \label{eq:int_split}
\end{equation}

\noindent
and contain the running coupling constant and the two leading,
logarithmically enhanced terms in the limit when $Q_0\ll Q$. The
two finite, non-logarithmic terms $-3/4$ and $-11/12$, respectively
emerge when integrating the non-singular part of the corresponding
splitting function in the limits $[0,\,1]$. Potentially, when $q/Q$
is not going to zero, these finite terms are larger than the
logarithmic terms and thus spoil an interpretation of the emerging
NLL-Sudakov form factor as a non-branching probability. Therefore,
without affecting the logarithmic order of the Sudakov form factors,
these finite terms are integrated over the interval $[q/Q,\,1-q/Q]$
rather than over $[q,\,Q]$. This way a Sudakov form factor determines
the probability for having no emission resolvable at scale $Q_0$
during the evolution from a higher scale $Q$ to a lower scale $Q_0$.
A ratio of two Sudakov form factors $\Delta(Q,Q_0)/\Delta(q,Q_0)$
then gives the probability for having no emission resolvable at
scale $Q_0$ during the evolution from $Q$ to $q$. Having
reweighted the matrix element, a smooth transition between this
and the parton shower region is achieved by choosing suitable
starting conditions for the shower evolution of the parton
ensemble and vetoing any parton shower emission that is harder than
the separation cut $k_{\perp,0}$.

\noindent
Within \sherpa\ the required matrix elements are provided by its
internal matrix element generator \amegic \cite{Krauss:2001iv}
and the parton shower phase is handled by \apacic
\cite{Kuhn:2000dk,Krauss:2005re}. Beyond the comparisons presented
here the \sherpa predictions for $W$+multijets have already been
validated and studied for Tevatron and LHC energies in
\cite{Krauss:2004bs,Krauss:2005nu}. Results for the production
of pairs of $W$-bosons have been presented in \cite{Gleisberg:2005qq}.

\subsection{The Dipole Cascade and CKKW}
\label{sec:dipole-cascade-ckkw}

The dipole model\cite{Gustafson:1988rq,Gustafson:1986db} as
implemented in the \ariadne program\cite{Lonnblad:1992tz} is based
around iterating $2\rightarrow 3$ partonic splitting instead of the
usual $1\rightarrow 2$ partonic splittings in a conventional parton
shower.  Gluon radiation is modeled as being radiated coherently from
a color--anticolor charged parton pair. This has the advantage of eg.\
including first order correction to the matrix elements for $e^+e^-\to
q\bar{q}$ in a natural way and it also automatically includes the
coherence effects modeled by angular ordering in conventional showers.
The process of quark antiquark production does not come in as
naturally, but can be added\cite{Andersson:1990ki}.  The emissions in
the dipole cascade is ordered according to invariant transverse
momentum defined as

\begin{equation}
   p_\perp^2 = \frac{s_{12}s_{23}}{s_{123}},\label{eq:arinpt}
\end{equation}
where $s_{ij}$ is the squared invariant mass of parton $i$ and $j$,
with the emitted parton having index 2.

When applied to hadronic collisions, the dipole model does not
separate between initial and final state radiation. Instead all
emissions are treated as coming from final state
dipoles\cite{Andersson:1989gp,Lonnblad:1996ex}.  To be able to extend
the dipole model to hadron collisions, extended colored objects are
introduced to model the hadron remnants.  Dipoles involving hadron
remnants are treated in a similar manner to the normal final-state
dipoles.  However, since the hadron remnant is considered to be an
extended object, emissions with small wavelength are suppressed.  This
is modeled by only letting a fraction of the remnant take part in the
emission. The fraction that is resolved during the emission is given
by
\begin{equation}
   a(p_\perp) = \left(\frac{\mu}{p_\perp}\right)^\alpha,\label{eq:arsup}
\end{equation}
where $\mu$ is the inverse size of the remnant and $\alpha$ is the
dimensionality.

There are two additional forms of emissions which need to be included
in the case of hadronic collisions.  One corresponds to an initial
state $g\rightarrow q \bar{q}$\cite{Lonnblad:1995wk}.  This does not
come in naturally in the dipole model, but is added by hand in a way
similar to that of a conventional initial-state parton
shower\cite{Lonnblad:1995wk}. The other corresponds to the
initial-state $q\to gq$ (with the gluon entering into the hard
sub-process) which could be added in a similar way, but this has not
been implemented in \ariadne yet.

When implementing CKKW for the dipole cascade, the procedure is
slightly different from what has been described
above\cite{Lonnblad:2001iq,Lavesson:2005xu}. First, rather than just
reconstructing emission scales using the $k_\perp$-algorithm, a
complete dipole shower history is constructed for each state produced
by the Matrix Element generator, basically answering the question
\textit{how would \ariadne have generated this state}. This will
produce a complete set of intermediate partonic states, $S_i$, and the
corresponding emission scales, $p_{\perp i}$.

The Sudakov form factors are then introduced using the Sudakov veto
algorithm.  The idea is that we want to reproduce the Sudakov form
factors used in Ariadne. This is done by performing a trial emission
starting from each intermediate state $S_i$ with $p_{\perp i}$ as a
starting scale. If the emitted parton has a $p_\perp$ higher than
$p_{\perp i+1}$ the state is rejected. This correspond to keeping the
state according to the no emission probability in Ariadne, which is
exactly the Sudakov form factor.

It should be noted that for initial-state showers, there are two
alternative ways of defining the Sudakov form factor. The definition
in eq.~(\ref{eq:sud}) is used in eg.\ \herwig\cite{Corcella:2000bw},
while eg.\ \pythia\cite{Sjostrand:2000wi,Sjostrand:2003wg} uses a form
which includes ratios of parton densities. Although formally
equivalent to leading logarithmic accuracy, only the latter
corresponds exactly to a no-emission probability, and this is the one
generated by the Sudakov-veto algorithm. This, however, also means
that the reconstructed emissions need not only be reweighted by the
running $\alpha_S$ as in the standard CKKW procedure above, but also
with ratios of parton densities, which in the case of gluon emissions
correspond to the suppression due to the extended remnants in
eq.~(\ref{eq:arsup}) as explained in more detail in
\cite{Lavesson:2005xu}, where the complete algorithm is presented.

\subsection{The MLM proceedure}
\label{sec:mlm-proceedure}
\def \et {\mbox{$E_T$}} \def \pt {\mbox{$p_T$}}
\def\etmin{\mbox{$E_T^{min}$}} In this approach we match the partons
from the ME calculation to the jets reconstructed after the
perturbative shower.  Parton-level events are defined by a minimum
\et\ threshold \etmin\ for the partons, and a minimum separation among
them, $\Delta R_{jj}>R_{min}$.  A tree structure is defined in analogy
with the CKKW algorithm, starting however from the colour-flow
extracted from the matrix-element
calculation~\cite{Caravaglios:1998yr}, thus defining the scales at
which the various powers of $\alpha_s$ are calculated. However, no
Sudakov reweighting is applied. Rather, events are showered, without
any hard-emission veto during the shower. After evolution, a jet cone
algorithm with cone size $R_{min}$ and minimum transverse energy
\etmin\ is applied to the final state. Starting from the hardest
parton, the jet which is closest to it in $(\eta,\phi)$ is selected.
If the distance between the parton and the jet centroid is smaller
than $R_{min}$, the parton and the jet match. The matched jet is
removed from the list of jets, and matching for subsequent partons is
performed. The event is fully matched if each parton has a matched
jet.  Events which do not match are rejected. A typical example is
when two partons are so close that they cannot generate independent
jets, and therefore cannot match. Rejection removes double counting of
the leading double logarithms associated to the collinear behaviour of
the amplitude when two partons get close.  Another example is when a
parton is too soft to generate its own jet, again failing matching.
This removes double counting of some single logarithms.  For events
which satisfy matching, it is furthermore required that no extra jet,
in addition to those matching the partons, be present.  Events with
extra jets are rejected, a suppression replacing the Sudakov
reweighting used in the CKKW approach. Events obtained by applying
this procedure to the parton level with increasing multiplicity can
then be combined to obtain fully inclusive samples spanning a large
multiplicity range. Events with extra jets are not rejected in the
case of the sample with highest partonic multiplicity.  The
distributions of observables measured on this inclusive data set
should not depend on the value of the parameters \etmin\ and
$R_{min}$, similar to the $k_{\perp,0}$ independence of the CKKW
approach. This algorithm is encoded in the \alpgen\ 
generator~\cite{Mangano:2001xp,Mangano:2002ea}, where evolution with
both \herwig\ and \pythia\ are enabled. In the following studies, the
results quoted as ``\alpgen'' employ the MLM matching scheme, and use
\alpgen for the generation of the parton-level matrix elements and
\herwig for the shower evolution and hadronisation.

\section{Examples and comparisons}
We present in this Section some concrete examples.  We concentrate on
the case of $W$+multijet production, which is one of the most studied
final states because of its important role as a background to top
quark studies at the Tevatron. At the LHC, $W$+jets, as well as the
similar $Z$+jets processes, will provide the main irreducible
backgrounds to signals such as multijet plus missing transverse
energy, typical of Supersymmetry and of other manifestations of new
physics. The understanding of $W$+multijet production at the Tevatron
is therefore an essential step towards the validation and tuning of
the tools presented here, prior to their utilization at the LHC.

For each of the three codes we calculated a large set of observables,
addressing inclusive properties of the events (\pt\ spectrum of the
$W$ and of leading jets), geometric correlations between the jets, and
intrinsic properties of the jets themselves, such as energy shapes.
In view of the limited space available here we present only a subset
of our studies, which will be documented in more detail in a future
publication.  An independent study of the systematics in the
implementation of the CKKW prescription in \herwig and \pythia was
documented in~\cite{Mrenna:2003if}.

The comparison between the respective results shows a reasonable
agreement among the three approaches, but points also to differences,
in absolute rates as well as in the shape of individual distributions,
which underscore the existence of an underlying systematic
uncertainty. The differences are nevertheless by and large consistent
with the intrinsic systematic uncertainties of each of the codes, such
as the dependence on the generation cuts or on the choice of
renormalization scale. There are also differences due to the choice of
parton cascade. In particular the \ariadne cascade is quite different
from a conventional parton shower, and it has been shown in this
workshop \cite{Kersevan:proc} that \ariadne eg.\ gives a much harder
$p_{\perp\mbox{W}}$ spectrum than does \herwig or \pythia. Now,
although the hard emissions in the matching proceedures should be
described by the exact matrix element, the Sudakov formfactors in the
\ariadne matching (and indirectly in the MLM scheme) are generated by
the cascade. In addition, the events in the \ariadne matching are
reweighted by PDF ratios in the same way as is done in the plain
cascade. This means that some properties of the cascade may affect
also the hard emissions in the matching procedure in these cases.

The existence in each of the codes of parameters specifying the
details of the matching algorithms presents therefore an opportunity
to tune each code so as to best describe the data. This tuning should
be seen as a prerequisite for a quantitative study of the overall
theoretical systematics: after the tuning is performed on a given set
of final states (e.g. the $W$+jets considered here), the systematics
for other observables or for the extrapolation to the LHC can be
obtained by comparing the difference in extrapolation between the
various codes. It is therefore auspicable that future analysis of
Tevatron data will provide us with spectra corrected for detector
effects in a fashion suitable to a direct comparison against
theoretical predictions.

The following two sections present results for the Tevatron
($p\bar{p}$ collisions at 1.96~TeV) and for the LHC ($pp$ at
14~TeV), considering events with a positively charged
$W$. Jets are defined by Paige's {\tt GETJET} cone-clustering
algorithm, with a calorimeter segmentation of ($\Delta \eta$,
$\Delta\phi$) = (0.1,$6^\circ$) and a cone size of 0.7 and 0.4 for
Tevatron and LHC, respectively. At the Tevatron (LHC) we consider jets
with $E_T>10(20)$~GeV, within $\vert \eta \vert < 2(4.5)$.
We use the PDF set CTEQ6L, with $\alpha_S(M_Z)=0.118$.

For our default distributions, the \alpgen\ results for the Tevatron
(LHC) were obtained using parton level cuts of $p_{T,min}=10(20)$ GeV,
$\vert \eta \vert<2.5(5)$, $R{jj}<0.7(0.4)$ and matching defined by
$E_{Tmin}=10$~GeV and $R=0.7$. The \sherpa\ samples have been
generated using matrix elements with up to four extra jets and the
value of the merging scale has been chosen to $k_{\perp,0} =
10(20)$~GeV, respectively. Finally, for \ariadne, the parton level
cuts were $p_{T,min}=10(20)$, $R{jj}<0.5(0.35)$ and, in addition, a
cut on the maximum pseudorapidity of jets, $\eta_{j\max}=2.5(5.0)$.

In all cases, the analysis is done at the hadron level, but without
including the underlying event.

\subsection{Tevatron Studies}
\label{sec:tevatron-results}
\begin{figure}
\begin{center}
\includegraphics[width=0.85\textwidth,clip]{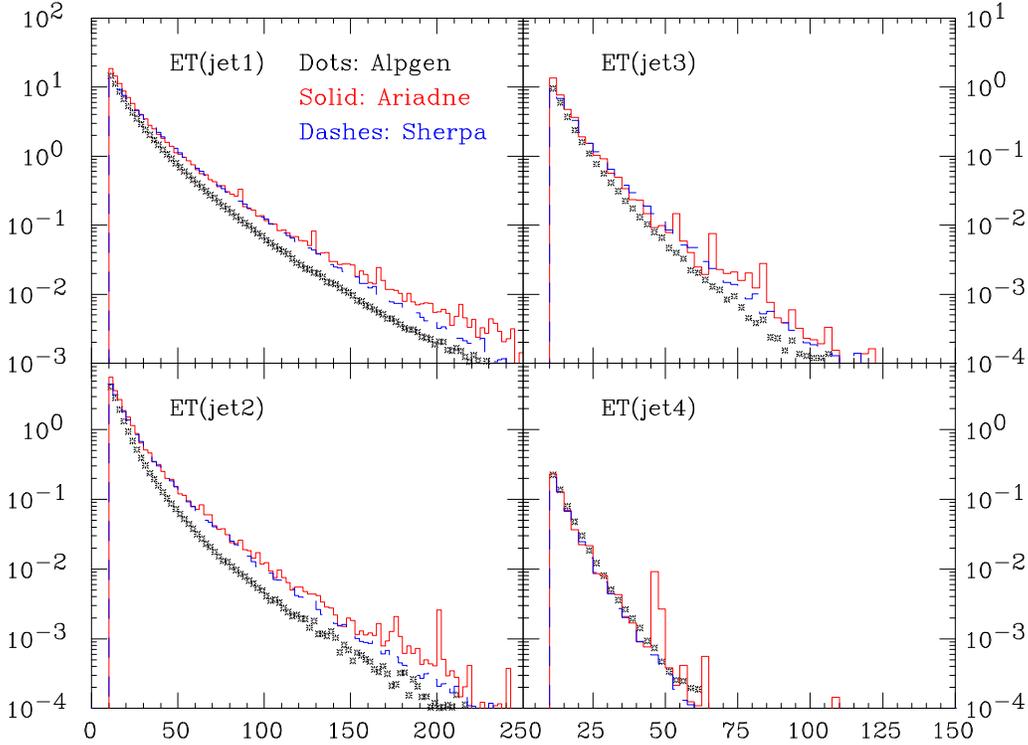}
\end{center}
\vskip -0.4cm

\caption{\label{fig:tev-pt}Inclusive $E_T$ spectra of the leading 4 jets at the Tevatron
(pb/GeV).}
\end{figure}

\begin{figure}
\begin{center}
\includegraphics[width=0.85\textwidth,clip]{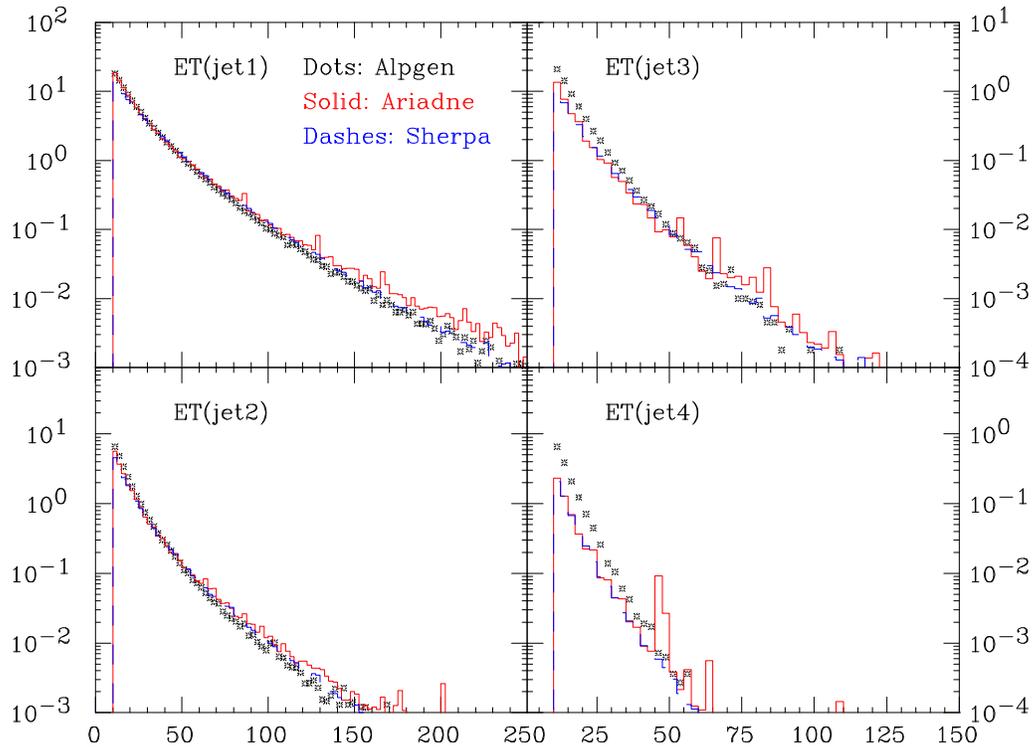}
\end{center}
\vskip -0.4cm

\caption{\label{fig:tev-pt05}Same as Fig.~1, but with the \protect\alpgen renormalization  
  scale reduced by a factor 2.}
\end{figure}

We start by showing in fig.~\ref{fig:tev-pt} the inclusive $E_T$
spectra of the leading 4 jets.  The absolute rate predicted by each
code is used, in units of pb/GeV.  We notice that the \alpgen spectrum
for the first two jets is softer than both \sherpa and \ariadne, with
the latter having even harder tails. The spectra for the third and
fourth jet are instead in very good agreement, both in shape and
normalization. As an indication of possible sources of systematics in
these calculations, we rescaled the renormalization scale used in
\alpgen by a factor of 1/2. As seen in fig.~\ref{fig:tev-pt05} the
distributions for the leading jets is now in perfect agreement with
\sherpa, with an increase in rate for the third and fourth jet. These
plots give us an idea of the level of flexibility which is intrinsic
in the calculation of higher-order jet production. One should not
forget that the rate for production of $N$ jets is proportional to the
$N$th power of $\alpha_s$, and the absence of the full set of virtual
corrections unavoidably leads to a large scale uncertainty.

\begin{figure}
\begin{center}
\includegraphics[width=0.85\textwidth,clip]{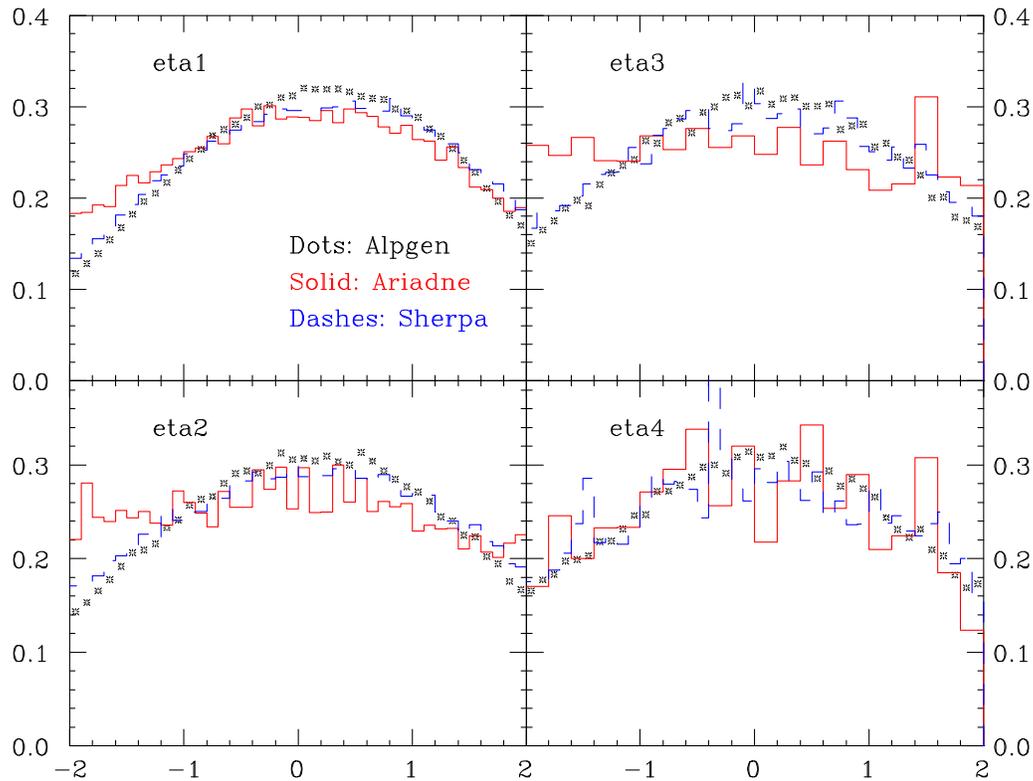}
\end{center}
\vskip -0.4cm
\caption{\label{fig:tev-eta}Inclusive $\eta$ spectra of the leading 4 jets at the
   Tevatron, normalized to unit area.
}

\end{figure}

Figure~\ref{fig:tev-eta} shows the inclusive $\eta$ spectra of the
leading 4 jets, all normalized to unit area.  The asymmetry for the
first two jets is due to the $W+$, which preferentially moves in the
direction of the proton (positive $\eta$). This is partially washed
out in the case of the third and fourth jet. There is a good agreement
between the spectra of \alpgen and \sherpa, while \ariadne spectra
appear to be broader, in particular for the subleading jets. This
broadening is expected since the gluon emissions in \ariadne are
essentially unordered in rapidity, which means that the Sudakov form
factors applied to the ME-generated states include also a $\log1/x$
resummation absent in the other programs.

\begin{figure}
\begin{center}
\includegraphics[width=0.85\textwidth,clip]{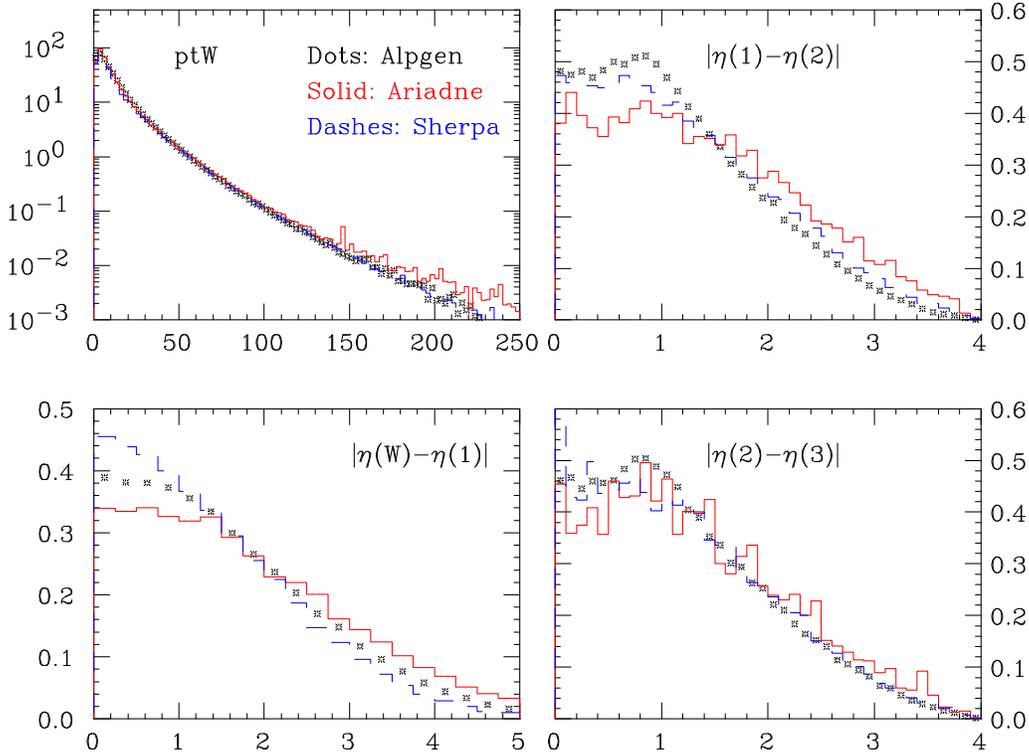}
\end{center}
\vskip -0.4cm
\caption{\label{fig:tev-ptw}Top left: inclusive $p_\perp(W^+)$ spectrum, pb/GeV. Bottom left:
  $|\eta(W^+) - \eta({\mathrm{jet}}_1)|$ (unit area). Top right:
  $|\eta({\mathrm{jet}}_1) - \eta({\mathrm{jet}}_2)|$ and bottom
  right: $|\eta({\mathrm{jet}}_2) - \eta({\mathrm{jet}}_3)|$ (unit
  area).}

\end{figure}

The top-left plot of fig.~\ref{fig:tev-ptw} shows the inclusive
$p_T$ distribution of the $W^+$ boson, with absolute normalization in
pb/GeV. This distribution reflects in part the behaviour observed for
the spectrum of the leading jet, with \alpgen slightly softer, and
\ariadne slightly harder than \sherpa.  The $\vert \eta \vert$
separation between the $W$ and the leading jet of the event is shown
in the top-right plot. The two lower plots show instead the
distributions of $|\eta({\mathrm{jet}}_1) - \eta({\mathrm{jet}}_2)|$
and $|\eta({\mathrm{jet}}_2) - \eta({\mathrm{jet}}_3)|$.  These last
three plots are normalized to unit area.  In all these cases, we
observe once more a reflection of the behaviour observed in the
inclusive $\eta$ distributions of the jets: \alpgen is slightly
narrower than \sherpa, and \ariadne is slightly broader.

\subsection{LHC Predictions}
\label{sec:lhc-predictions}
In this section we confine ourselves to \alpgen and \sherpa. It turns
out that \ariadne has a problem in the reweighting related to the fact
that initial-state $g\to q\bar{q}$ emissions, contrary to the gluon
emissions, are ordered both in $p_\perp$ and rapidity. With the
extra phase space available at the LHC this leads to unnatural
reconstructions which, in turn, gives rise to a systematically too
high reweighting. A solution for this problem is under investigation
and a fuller comparison including \ariadne will be documented in a
future publication.

\begin{figure}
\begin{center}
\includegraphics[width=0.85\textwidth,clip]{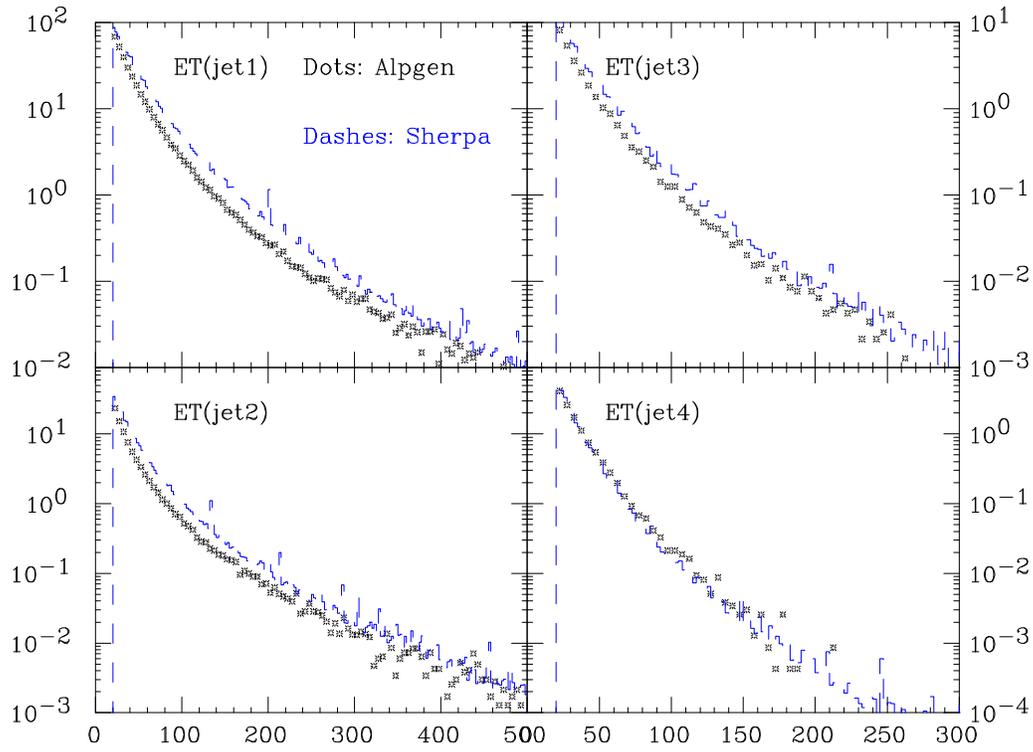}
\end{center}
\vskip -0.4cm

\caption{\label{fig:lhc-pt}Inclusive $E_T$ spectra of the leading 4 jets at the LHC
(pb/GeV).}
\end{figure}

Following the same sequence of the Tevatron study, we start by showing
in fig.~\ref{fig:lhc-pt} the inclusive $E_T$ spectra of the leading 4
jets.  The absolute rate predicted by each code is used, in units of
pb/GeV. The relative behaviour of the predictions by \alpgen and
\sherpa follows the pattern observed in the Tevatron case, with
\alpgen being softer in the case of the leading two jets. We do not
notice however a deterioration of the discrepancy going from the
Tevatron to the LHC, suggesting that once a proper tuning is achieved
at lower energy the predictions of two codes for the LHC should be
comparable.

\begin{figure}
\begin{center}
\includegraphics[width=0.85\textwidth,clip]{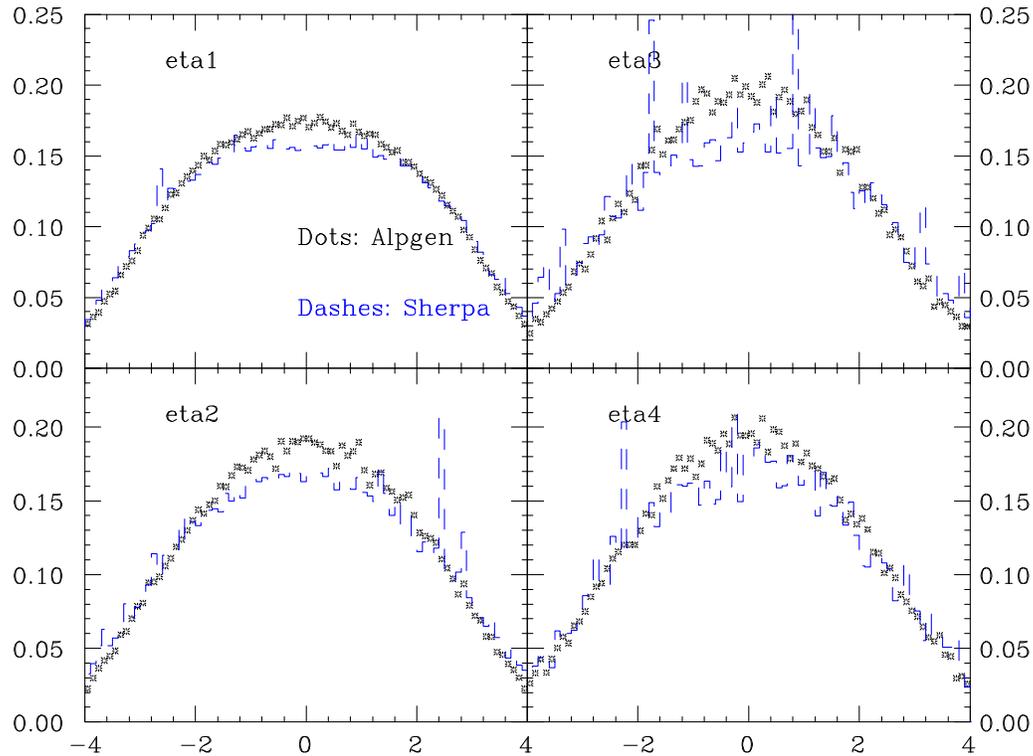}
\end{center}
\vskip -0.4cm
\caption{\label{fig:lhc-eta}Inclusive $\eta$ spectra of the leading 4 jets at the
   LHC, normalized to unit area.
}

\end{figure}

Figure~\ref{fig:lhc-eta} shows the inclusive $\eta$ spectra of the leading 4
jets, all normalized to unit area.
The asymmetry now is not present, because of the symmetric rapidity
distribution of the $W^+$ in $pp$ collisions.
As in the case of the Tevatorn, jet production in \alpgen is slightly
more central than in \sherpa.

\begin{figure}
\begin{center}
\includegraphics[width=0.85\textwidth,clip]{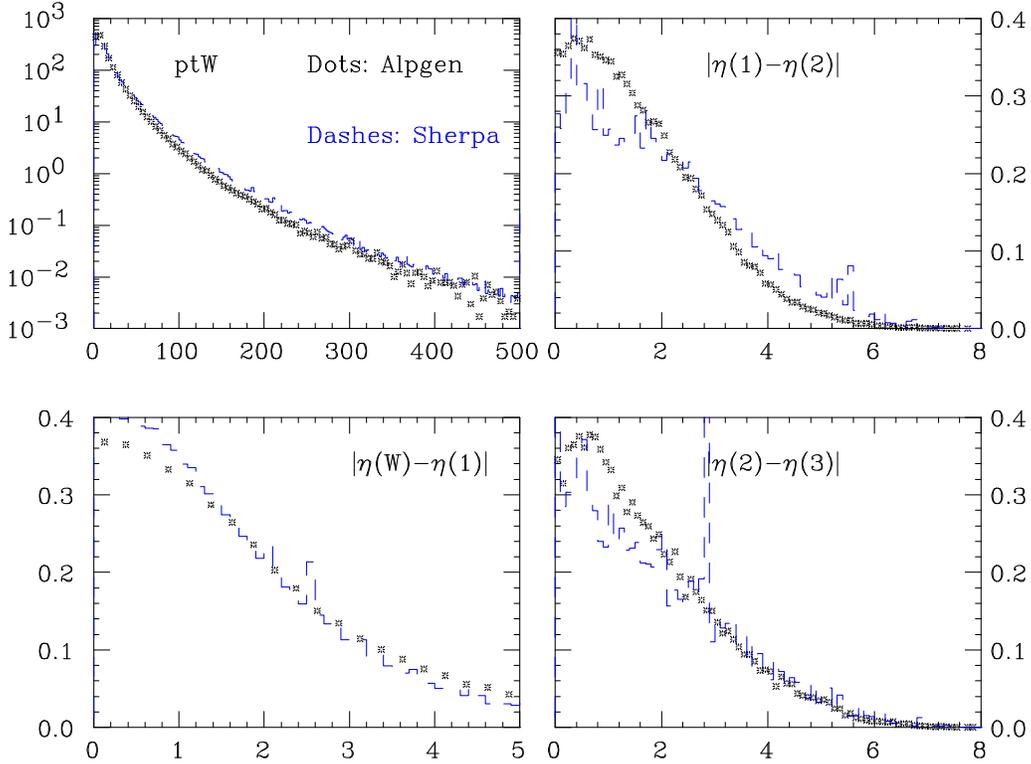}
\end{center}
\vskip -0.4cm
\caption{\label{fig:lhc-ptw}Top left: inclusive $pt_T(W^+)$ spectrum, pb/GeV. Bottom left:
   $|\eta(W^+) - \eta({\mathrm{jet}}_1)|$ (unit area). Top right:
$|\eta({\mathrm{jet}}_1) - \eta({\mathrm{jet}}_2)|$
and bottom right: $|\eta({\mathrm{jet}}_2) - \eta({\mathrm{jet}}_3)|$ (unit area).}

\end{figure}

The top-left plot of fig.~\ref{fig:lhc-ptw} shows the
inclusive $p_T$ distribution of the $W^+$ boson, with absolute
normalization in pb/GeV. The $\vert \eta \vert$
separation between the $W$ and the leading jet of the event is shown
in the top-right plot. The two lower plots show instead the
distributions of $|\eta({\mathrm{jet}}_1) - \eta({\mathrm{jet}}_2)|$
and $|\eta({\mathrm{jet}}_2) - \eta({\mathrm{jet}}_3)|$.
These last three plots are normalized to unit area.
As before, the features of these comparisons reflect what observed in
the inclusive jet properties.

\section{Conclusions}
\label{sec:conclusions}
This document summarizes our study of a preliminary comparison of
three independent approaches to the problems of merging matrix element
and parton shower evolution for multijet final states.  Overall, the
picture shows a general consistency between the three approaches,
although there are occasional differences.  The origin of these
differences is under study. It could be based on intrinsic differences
between the matching schemes, as well as to differences between the
different shower algorithms used in the three cases. We expect
nevertheless that these differences be reconciled with appropriate
changes in the default parameter settings for the matching schemes, as
partly supported by the few systematic studies presented
here. Validation and tuning on current Tevatron data is essential, and
will allow to reduce the systematics.

It is also important to compare these models to HERA data. However,
besides some preliminary investigations for \ariadne
\cite{abergthesis}, there is no program which properly implement a
CKKW or MLM matching scheme for DIS. The energy of HERA is, of course,
lower, as are the jet multiplicities and jet energies, but HERA has
the advantage of providing a large phase space for jet production
which is not mainly determined by the hard scale, $Q^2$, but rather by
the total energy, giving rise to large logarithms of $x\approx
Q^2/W^2$ which need to be resummed to all orders. This is in contrast
to the Tevatron, where the phase space for additional jets in
W-production mainly are determined by $m_W$. However, when going to
the LHC there may also be important effects of the increased energy,
and there will be large logarithms of $x\propto m_W/\sqrt{S}$ present,
which may need to be resummed. The peculiar treatment of the available
phase space in the plain \ariadne cascade means that some logarithms
of $x$ are resummed in contrast to conventional initial-state parton
cascades.  This feature survives the matching procedure and is the
reason for the broader rapidity spectra presented in the figures
above. In DIS this is reflected by the increased rate of forward jets,
and such measurements are known to be well reproduced by \ariadne
while conventional parton showers fail. It would be very interesting
if the matching of these conventional showers with higher order matrix
elements would improve the description of forward jets. In that case
the extrapolation of the Tevatron results to the LHC would be on much
safer grounds.

As our study of the LHC distributions suggests, the increase in energy
exhibits the same pattern of discrepancies observed at the Tevatron.
We therefore expect that if different algorithms are tuned on the same
set of data, say Tevatron $W$+jets, they will extrapolate in the same
way to the LHC or to different final states, for example multijet
configurations without $W$ bosons.  While these systematics studies
can be performed directly at the Monte Carlo level, only the
availability of real measurements from the Tevatron can inject the
necessary level or realism in these exploration. We look forward to
the availability of such data.
\bibliographystyle{heralhc}
{\raggedright
\bibliography{meps}
}
\end{document}